\documentclass[twocolumn,tighten,trackchanges]{aastex701}
\usepackage{CJK}
\usepackage{amsmath}
\usepackage{multirow}

\newcommand{\hostsub}{\texttt{HostSub\_GP}}
\newcommand{\pypeit}{\texttt{PypeIt}}

\graphicspath{{./}{figures/}}

\begin{document}
\begin{CJK*}{UTF8}{gbsn}
\title{\texttt{HostSub\_GP}: Precise Galaxy Background Subtraction in Transient {Long-slit} Spectroscopy with Gaussian Processes}

\author[0000-0002-7866-4531]{Chang~Liu (刘畅)}
\affil{Department of Physics and Astronomy, Northwestern University, 2145 Sheridan Rd, Evanston, IL 60208, USA}
\affil{Center for Interdisciplinary Exploration and Research in Astrophysics (CIERA), Northwestern University, 1800 Sherman Ave, Evanston, IL 60201, USA}
\affil{NSF-Simons AI Institute for the Sky (SkAI), 172 E. Chestnut St., Chicago, IL 60611, USA}
\email{ptg.cliu@u.northwestern.edu}

\author[0000-0001-9515-478X]{Adam~A.~Miller}
\affil{Department of Physics and Astronomy, Northwestern University, 2145 Sheridan Rd, Evanston, IL 60208, USA}
\affil{Center for Interdisciplinary Exploration and Research in Astrophysics (CIERA), Northwestern University, 1800 Sherman Ave, Evanston, IL 60201, USA}
\affil{NSF-Simons AI Institute for the Sky (SkAI), 172 E. Chestnut St., Chicago, IL 60611, USA}
\email{amiller@northwestern.edu}

\correspondingauthor{Chang Liu}

\begin{abstract}

We present a novel host galaxy subtraction technique in long-slit spectroscopy for extragalactic transients. Unlike classic methods which generally estimate the background using a simple linear interpolation of local galaxy flux in the 2D spectrum, our approach leverages multi-band archival images of the host galaxies to model the background emission from the galaxy in the 2D spectrum. Such imaging {encodes} the wavelength-dependent galaxy profile along the slit{, and} is readily accessible through wide-field imaging surveys. We construct a smooth prior for the 2D galaxy profile with a Gaussian process (GP) based on these reference images, and use another GP to model the correlated deviations from the prior in the observed spectrum. This enables accurate inference of the galaxy flux blended with the transient. On synthetic long-slit data of a spiral galaxy extracted from a Multi Unit Spectroscopic Explorer hyper-spectral cube, the GP method remains robust as long as the host galaxy is spatially resolved and consistently outperforms  {classic methods}. We apply the method to archival Keck spectra of two real transients, SN\,2019eix and AT\,2019qiz, to further demonstrate how the method uniquely recovers weak spectral features amid strong galaxy contamination, enabling refined constraints on the properties of both transients. We have released the software implementation, \hostsub, a scalable toolkit that leverages \texttt{JAX}, with an MIT license.

\end{abstract}



\section{Introduction} \label{sec:intro}

Deblending the photons of extragalactic transients from their host galaxies is a fundamental challenge in observational time-domain astronomy. Precise host subtraction in spectroscopy is critical in both classifying new transients and follow-up studies. Host contamination dilutes the spectral features from the transients, making robust classification particularly difficult for transients overlapping a high background (e.g., galaxy nuclei). For example, among the Type Ia supernovae (SNe\,Ia) discovered by the Zwicky Transient Facility, ambiguous subtyping is prevailing among those near the galaxy cores \citep{Dimitriadis_ZTF_2025}. A biased classification rate towards high background regions adds complexity to the rate studies on certain transient types. While fitting the joint spectral energy distribution (SED) of the transient and its host \citep[e.g.,][]{SF_2005, NGSF_2022, DASH_2019} improves the classification accuracy on host contaminated spectra \citep{Milligan_2025}, these methods still struggle when the galaxy light dominates.
Host contamination also impacts {the quantitative measurement} of spectral features. Even for SNe near maximum luminosity which outshine their background, host contamination {can bias} line velocity measurements \citep{Burgaz_2025}. At later phases, when SNe are generally faint compared to their background, the effects of host contamination {become even more pronounced}.
Many existing spectroscopic samples of late-time SNe are limited to objects with large offsets from their host galaxies \citep[e.g.,][]{Graham_2017,Maguire_2018}, which introduces substantial selection biases in the local host environment. Other samples inevitably include objects with significant residual galaxy contamination \citep[e.g.,][]{Maguire_IIP_2012,Tucker_2020,Fang_2022,Das_2023}, which compromise quantitative inferences of the SN properties, such progenitor masses of both SNe\,Ia \citep{Flors_limits_2018, Flors_2020} and core-collapse SNe \citep{Dessart_2020, Fang_red_2025, Fang_conciling_2025}.

Precise transient flux measurements in photometry often leverage archival reference images of the host galaxies. As a routine in modern time-domain surveys, image subtraction techniques \citep[e.g.,][]{Alard_1998}, which match the point spread functions (PSFs) between the new and the reference images to robustly subtract the background flux, play a vital role in transient discovery and producing reliable flux measurements \citep[e.g.,][]{ASASSN_2014, ZTF_data_2019}. In contrast, direct image subtraction is far more difficult in spectroscopy: (i) pre-transient archival spectroscopic observations suitable for subtraction rarely, if ever, exist; (ii) post-transient spectroscopic reference observations are unavailable until the transient has fully faded, and are significantly more resource-intensive (e.g., requiring longer exposure times on larger telescopes) than photometric references; (iii) slit losses (due to the wavelength dependence of the seeing) and atmospheric extinction are typically different when obtaining reference spectra and (iv) matching PSFs between new and reference observations is tricky -- while PSF reconstruction on hyper-spectral data obtained with integral field spectrographs (IFS), which preserve two-dimensional (2D) spatial information, is feasible \citep{Courbin_2000, Bongard_2011, HyperGal_2022}, it becomes extremely challenging in long-slit spectroscopy, the most widely used technique, where one of the two spatial dimensions is inherently lost.

Classic long-slit data reduction process usually treats transients as isolated sources. The background -- comprising both sky emission (e.g., atmospheric emission, solar system and extragalactic background) and host galaxy light -- is assumed to be smooth near the transient with little fluctuation and estimated by interpolating the flux measured beside the transient's trace in the 2D spectrum, using regions outside the trace that are selected around it. However, the smoothness assumption no longer holds when the host environments have steep light profiles (e.g., galactic nuclei or bulges) or complex substructures (e.g., spiral arms or clumpy star-forming regions), common for extragalactic transients like SNe. Depending on the local light profile and regions chosen for background estimation, the classic technique can lead to either over- or under-subtraction of the host galaxy light. This introduces unpredictable systematic biases into the extracted transient spectrum, which cannot be reliably captured by standard uncertainty estimates. More advanced techniques can provide cleaner host galaxy subtraction in long-slit spectroscopy, but they typically require either firm knowledge of PSFs \citep[e.g.,][]{Lucy_2003, Blondin_2005}, or concurrent multi-band photometry with the spectrum, as well as host galaxy SED models \citep[e.g.,][]{Ellis_2008, Foley_2008}. In practice, these requirements are often difficult to meet.

Motivated by the success of image subtraction technique in photometry, we present a novel approach for host subtraction in long-slit spectroscopy. It utilizes archival images from all-sky imaging surveys that span the entire optical wavelength range, including the Sloan Digital Sky Survey \citep[SDSS;][$ugriz$]{SDSS_2000}, the Panoramic Survey Telescope and Rapid Response System \citep[Pan-STARRS, or PS;][$grizy$]{PanSTARRS_2016}, the Dark Energy Spectroscopic Instrument (DESI) Legacy Imaging Surveys \citep[LS;][$griz$]{LS_2019}, and in the near future, the Legacy Survey of Space and Time \citep[LSST;][$ugrizy$]{LSST_2019}. From these multi-band images, we can construct a reasonable prior of the galaxy profile in a data-driven way via Gaussian processes (GPs) to avoid expensive SED fitting. Our software implementation, \hostsub, is a scalable toolkit built upon \texttt{tinygp} \citep{tinygp_2024}, leveraging the just-in-time compilation and automatic differentiation in \texttt{JAX} \citep{jax_2018}. The open source software is available at \url{https://github.com/slowdivePTG/HostSub_GP} with an MIT license.

In Section~\ref{sec:method}, we introduce the methodology of 2D galaxy spectrum modeling with GPs, followed by an elaboration of the pipeline modules in Section~\ref{sec:pipeline}. We show that the GP method significantly outperforms the classic method {based on linear interpolation} in Section~\ref{sec:result} by testing it on a synthetic dataset (Section~\ref{sec:result_syn}) and two real transients (Sections~\ref{sec:result_19eix} and \ref{sec:result_19qiz}). In Section~\ref{sec:limit}, we summarize the recommended use cases for applying \hostsub\ following the main assumptions we have made, before presenting conclusions in Section~\ref{sec:conclusion}.

\section{Methodology: Modeling the 2D Spectrum with GPs}\label{sec:method}
The ultimate goal of \hostsub\ is to estimate the galaxy light at each pixel of a 2D spectrum. The observed flux\footnote{{Throughout this paper, uppercase $F$ denotes the flux in a 1D spectrum, while lowercase $f$ denotes the flux in a 2D spectrum (spectrogram).}} 
$f_\mathrm{obs}(x, \lambda)$ is the function of a spatial coordinate $x$ and the wavelength $\lambda$ of each pixel, which consists of the contribution from the source $f_\mathrm{src}(x, \lambda)$, its host galaxy $f_\mathrm{host}(x, \lambda)$, and the sky background $f_\mathrm{sky}(\lambda)$. Here we assume a constant sky background along the slit (typically~$\lesssim$1\arcmin\ long). In a 2D spectrum with flat, bias, and dark fields calibrated, $f_\mathrm{sky}$ is only dependent on wavelength. {We also assume that the atmospheric differential refraction has been corrected during the observation, either by observing at the parallactic angle or using an Atmospheric Dispersion Corrector.}

The transient{, as a point source, contributes negligibly beyond a few PSF widths from its center}. To estimate $f_\mathrm{host} + f_\mathrm{sky}$, we mask the {transient's trace} and model the host and sky contributions {in the surrounding regions. This allows us} to infer the background flux blended with the transient.

Inspired by \citet{Lucy_2003}, {$f_\mathrm{host}$ could be modeled} as two semi-separable parts,
\begin{equation}\label{eq:host_sep}
    f_\mathrm{host}(x, \lambda) = F_\mathrm{host}(\lambda)\cdot \xi_\mathrm{host}(x, \lambda).
\end{equation}
Here, $F_\mathrm{host}(\lambda)$, which is effectively a 1D spectrum of the galaxy, only has wavelength dependence, and $\xi_\mathrm{host}(x,\lambda)$ is a normalized {observed} spatial profile{, i.e., the intrinsic galaxy profile convolved with the PSF}. 
We do not require the galaxy background to be homogeneous, i.e., $\xi_\mathrm{host}$ can be wavelength dependent, as the stellar population could be different across the region covered by the slit. Nevertheless, we expect $\xi_\mathrm{host}$ to vary slowly along the wavelength axis, whereas  $F_\mathrm{host}$ can vary on shorter scales (typically the instrumental resolution). Hence we need orders-of-magnitude fewer data points along the wavelength direction to model $\xi_\mathrm{host}$. In addition, multi-band pre-transient imaging from archival surveys provides critical prior knowledge of $\xi_\mathrm{host}$ that we will integrate into our model.\footnote{The {resolving power} of archival multi-band imaging is too low to serve as a useful prior for $F_\mathrm{sub}$ without SED modeling.}

In practice, {neither $f_\mathrm{host}$ nor $\xi_\mathrm{host}$} is directly measurable from the 2D spectrum{, which} is also blended with the sky background. To remove the sky emission, we define the global-sky-subtracted 2D spectrum $f_\mathrm{sub}$ as
\begin{equation}\label{eq:sub_sep}
    \begin{aligned}
        f_\mathrm{sub}(x,\lambda) &\equiv f_\mathrm{obs}(x,\lambda) - \langle f_\mathrm{obs}\rangle_{\mathrm{sky}}\\
    & = f_\mathrm{src}(x, \lambda) + F_\mathrm{sub}(\lambda)\cdot\xi_\mathrm{sub}(x,\lambda),
    \end{aligned}
\end{equation}
where $f_\mathrm{src}$ is considered negligible outside the trace mask, and
\begin{equation}\label{eq:xi_sub}
    \xi_\mathrm{sub}(x,\lambda)\equiv\xi_\mathrm{host}(x,\lambda) - \langle \xi_\mathrm{host}\rangle_{\mathrm{sky}}(\lambda).
\end{equation}
Here $\langle \cdot\rangle_\mathrm{sky}$ denotes averaging the 2D quantity along the spatial axis over {what we define as the global sky region with a width of $L_\mathrm{sky}$, e.g., for $\xi_\mathrm{host}$}
\begin{equation}
    \langle \xi_\mathrm{host}\rangle_{\mathrm{sky}}(\lambda) = \frac{1}{L_\mathrm{sky}}\int_\mathrm{sky} \xi_\mathrm{host}(x,\lambda)\mathrm dx,
\end{equation}
resulting in a function of wavelength.
Instead of modeling $\xi_\mathrm{host}$ directly, we estimate $\xi_\mathrm{sub}$. For simplicity, we force $\xi_\mathrm{sub}$ to be normalized over what we define as the host region,
\begin{equation}\label{eq:xi_sub_norm}
    \int_\mathrm{host} \xi_\mathrm{sub}(x,\lambda)\mathrm{d}x = 1,
\end{equation}
so the physical significance of $F_\mathrm{sub}$ is the 1D spectrum extracted from a specific host region on the sky-subtracted 2D spectrum $f_\mathrm{sub}$. It can be either positive or negative, depending on the relative brightness of the host and the sky regions. The selection of the two regions is very flexible, as long as (i) both regions are outside the trace mask; (ii) the average flux in the sky region is significantly different (either higher or lower) than the host region at all wavelengths{; otherwise, $\xi_\mathrm{sub}$ will diverge when $F_\mathrm{sub}$ approaches 0 at any wavelength}.

To bridge the archival images with the spectroscopic observation, we model both $F_\mathrm{sub}$ and $\xi_\mathrm{sub}$ with GPs, and implement the prior derived from the archival images as the mean function. In practice, we build a smooth model for $F_\mathrm{sub}$ using a 1D GP (notations with hats denote estimators),
\begin{equation}\label{eq:1D_GP}
    \widehat F_\mathrm{sub}(\lambda)\sim \mathcal{GP}(\mu_\mathrm{1D}, K_1(\lambda,\lambda^\ast)),
\end{equation}
which is conditioned on the observed values of $F_\mathrm{sub}$. Here $\mu_\mathrm{1D}$ is a mean value of the GP and $K_1$ is the kernel which determines the covariance between different wavelengths. Both $\mu_\mathrm{1D}$ and the {kernel} parameters of $K_1$ are hyperparameters of the GP model. 
Separately, we build a smooth model for $\xi_\mathrm{sub}$ with a 2D GP,
\begin{equation}\label{eq:2D_GP}
    \widehat \xi_\mathrm{sub}(x,\lambda)\sim\mathcal{GP}(\xi_\mathrm{sub,img}(x,\lambda), K_2((x,\lambda),(x^\ast,\lambda^\ast))),
\end{equation}
which is conditioned on the {measured} $\xi_\mathrm{sub}$ in the host region. Unlike the 1D GP which adopts a single, constant mean function $\mu_\mathrm{1D}$, the mean function of the 2D GP is derived from the (multi-band) archival images (see Section~\ref{sec:pipeline_prior}), which is a function of both the spatial coordinates and wavelength. This mean function serves as a reasonable prior of the galaxy light profile along the slit, and, as we demonstrate below, will dramatically improve the precision in estimating the host galaxy background. $K_2$ is the corresponding kernel which determines the covariance between ($x$, $\lambda$) pairs.

Finally, we derive a model of $f_\mathrm{sub}$ as the product of $\widehat F_\mathrm{sub}$ and $\widehat\xi_\mathrm{sub}$, so the estimated 2D source flux is simply
\begin{equation}\label{eq:f_sub_model}
    \widehat f_\mathrm{sub}(x,\lambda) = \widehat F_\mathrm{sub}(\lambda)\cdot\widehat\xi_\mathrm{sub}(x,\lambda).
\end{equation}
Comparing $\widehat f_\mathrm{sub}$ to $f_\mathrm{sub}$ yields the model likelihood, and the hyperparameters of the GP ($\mu_\mathrm{1D}$ and kernel parameters) are optimized by maximizing the posterior. {The implementation details of the methodology are elaborated in Section~\ref{sec:pipeline_main}.} 

The output of \hostsub\ is the host galaxy subtracted 2D spectrum,
\begin{equation}
    \widehat f_\mathrm{src}(x,\lambda) = f_\mathrm{sub}(x,\lambda) - \widehat f_\mathrm{sub}(x,\lambda).
\end{equation}
The 1D spectrum can be extracted from the 2D trace using existing pipelines, such as \texttt{iraf} \citep{IRAF_1986} and \pypeit\ \citep{pypeit:joss_pub}.

\section{Pipeline Overview}\label{sec:pipeline}

\begin{figure*}
    \centering
    \includegraphics[width=0.65\linewidth]{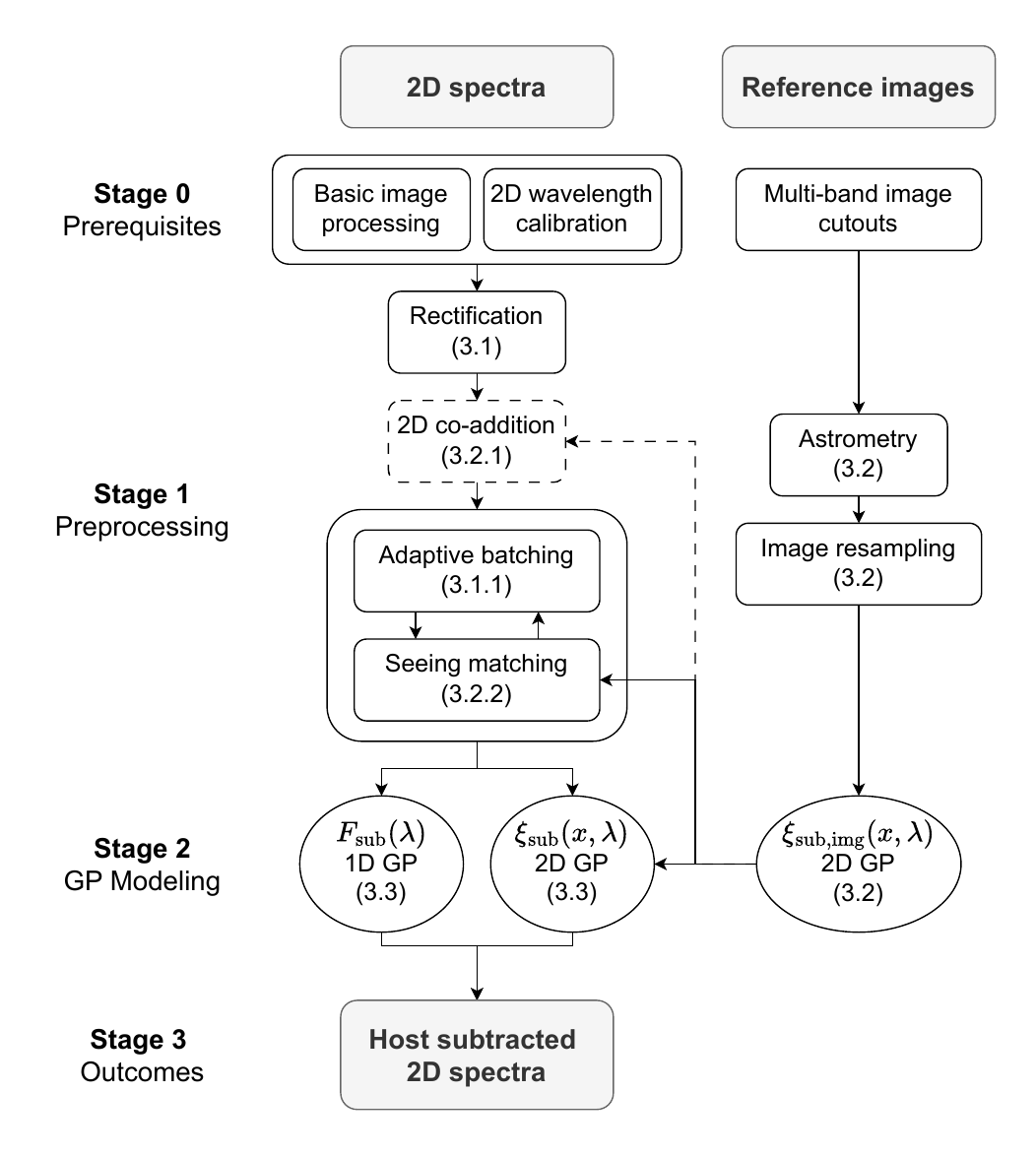}
    \caption{The data reduction framework in \hostsub. The 2D co-addition will only be performed if multiple exposures exist.}
    \label{fig:pipeline}
\end{figure*}

In this section, we provide an overview of the \hostsub\ framework and, for each of the modules, we elaborate the details of the data reduction and modeling procedures. 
Figure~\ref{fig:pipeline} is a high-level summary of the pipeline.

\subsection{Preprocessing}

\begin{figure*}
    \centering
    \includegraphics[width=\linewidth]{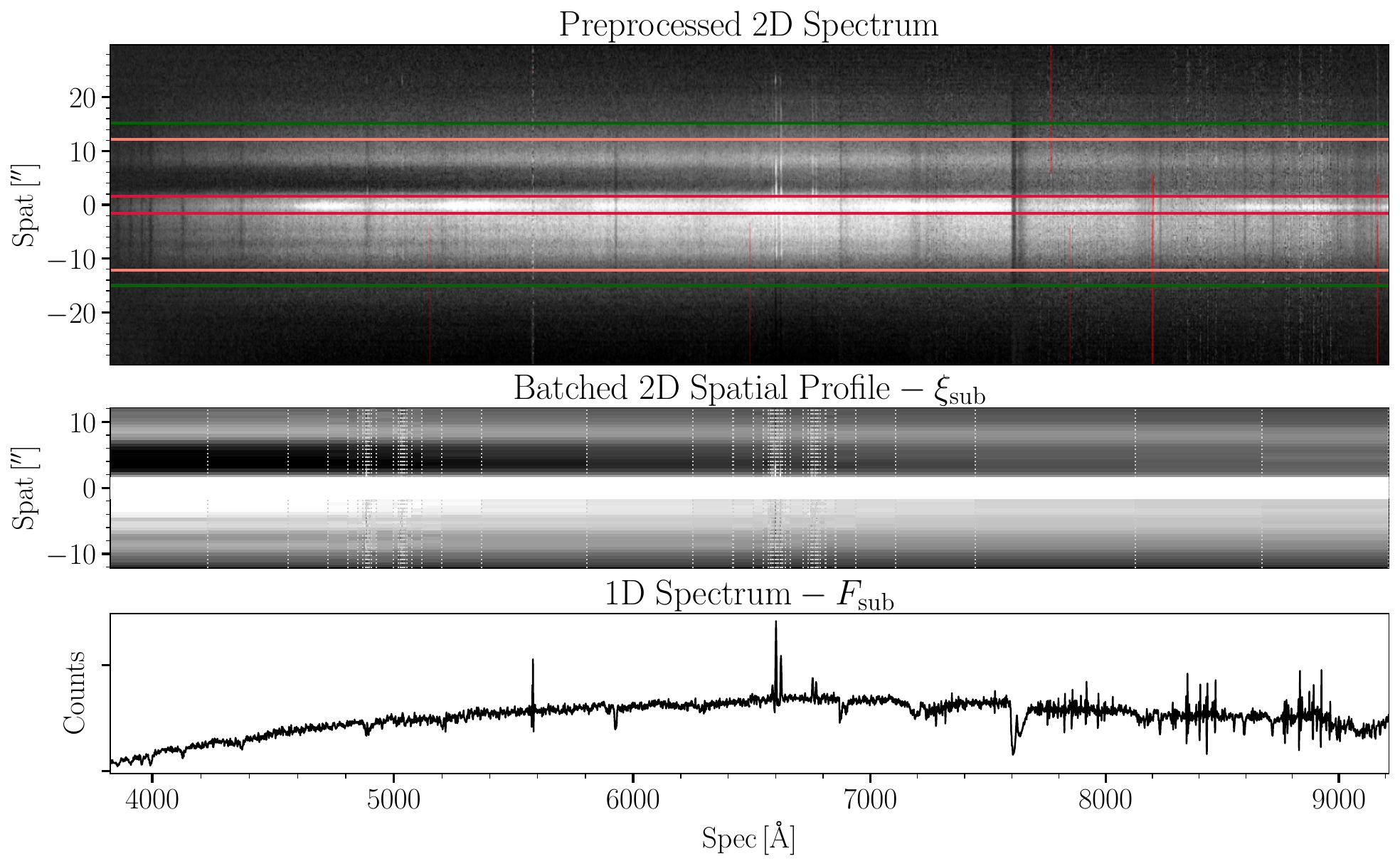}
    \caption{An example of the derived $F_\mathrm{sub}$ and $\xi_\mathrm{sub}$ from the preprocessed data {with adaptive binning around galaxy emission lines}. \textit{Upper panel:} the example preprocessed 2D spectrum after co-addition, rectification, and global background subtraction. The edges of the sky regions (green), host regions (salmon), and transient mask (red) are displayed. The bad pixels are indicated with red colors. \textit{Middle panel:} the batched 2D galaxy profile $\xi_\mathrm{sub}$. Dashed gray lines indicate batch edges. \textit{Bottom panel:} the 1D galaxy spectrum $F_\mathrm{sub}$ used to normalize the galaxy profile.}
    \label{fig:input}
\end{figure*}

As input \hostsub\ requires a fully calibrated 2D spectrum and a configuration file containing user-defined parameters. Basic image calibrations, including the flat, bias, and dark field corrections, bad pixel rejection, cosmic ray (CR) removal, and 2D wavelength solution (i.e., the corresponding wavelength at each given pixel in the 2D spectrum), should be done in the upstream data reduction pipeline (e.g., \pypeit). Flux calibration is not a requirement, because \hostsub\ does not model the SED of the host galaxies. 

Additionally, it is expected that a first round of object finding and tracing has been carried out in the upstream pipeline, so \hostsub\ can locate the trace of the transient at each wavelength $(x_\mathrm{tr}(\lambda), y_\mathrm{tr}(\lambda))$.
Here $x$ and $y$ roughly correspond to the spatial and spectral axes of the 2D spectrum, although the actual coordinate system is slightly curved. The estimate can be further fine tuned by aligning the galaxy light in the spectrum to the archival galaxy images, as detailed in Section~\ref{sec:pipeline_co-add}. 
The pipeline then calculates the sky separation from each pixel to the trace at the corresponding wavelength. 
Finally, the calibrated 2D spectrum (both the flux and the inverse variance of the measurements) is mapped onto a rectified grid of spatial (the sky separation from the transient in arcsecond) and spectral coordinates (the wavelength in Angstrom), to eliminate the curvature of the trace. 

On the rectified spectrum, {as illustrated in Section~\ref{sec:method},} users will need to {specify} three regions, each corresponding to a range of spatial distances to the trace of the transient: (i) a central mask, which contains the transient and is excluded in the modeling; (ii) the sky region, which is used to calculate the global sky $\langle f_\mathrm{obs}\rangle_{\mathrm{sky}}$ for each wavelength; and (iii) the host galaxy region, which is used to derive the 1D quantity $F_\mathrm{sub}$ and the 2D quantity $\xi_\mathrm{sub}$ that \hostsub\ actually models. Following rectification, each row of pixels corresponds to a fixed wavelength $\lambda_j$, so $\langle f_\mathrm{obs}\rangle_{\mathrm{sky}}(\lambda_j)$ is calculated as the mean flux in the sky region. After subtracting $\langle f_\mathrm{obs}\rangle_{\mathrm{sky}}$, $F_\mathrm{sub}(\lambda_j)$ can be calculated as the sum of the flux in the row within the host galaxy region. Eventually, for every individual pixel $i$ in the host galaxy region, $\xi_\mathrm{sub}(x_i, \lambda_j)$ is simply the ratio of $f_\mathrm{sub}(x_i,\lambda_j)$ and $F_\mathrm{sub}(\lambda_j)$, as defined in Eq~(\ref{eq:xi_sub}) and (\ref{eq:xi_sub_norm}). An example of the 2D spectrum preprocessed by \hostsub\ is displayed in Figure~\ref{fig:input}.

Once $F_\mathrm{sub}$ and $\xi_\mathrm{sub}$ are separated, we note again that $\xi_\mathrm{sub}$ typically varies slowly along the spectral direction, i.e., the scale length is much greater than the instrumental spectral resolution. By default, \hostsub\ combines the pixels along the spectral direction by a factor of 64 (which can be modified by users in the configuration file) before $\xi_\mathrm{sub}$ is evaluated. The coarse binning (bin widths $\gtrsim100$\,\r{A}) enhances the signal-to-noise ratio, while preserving the essential information, as the slow spectral variation of $\xi_\mathrm{sub}$ ensures that no significant details are lost.

\subsubsection{Galaxy Emission Line Searching and Adaptive Batching}\label{sec:pipeline_host_lines}

A notable exception where $\xi_\mathrm{sub}$ can fluctuate rapidly on the scale of the instrumental resolution occurs near the nebular emission lines of star-forming galaxies. The emission lines trace the youngest and most massive stars, whereas the spectral continuum and most absorption lines are shaped predominantly by the older stellar population. 
Since {younger and older populations often show different distributions in galaxies}, 
the spatial distribution of flux within the wavelength range of strong emission lines can be fairly different from the surrounding continuum. As a result, the emission lines require finer wavelength bins, whose widths correspond to the resolution of the spectrograph.

\hostsub\ automatically searches for prominent host galaxy emission lines and adjusts the bin size in an adaptive way. A candidate emission line will be labeled at wavelength $\lambda_j$ if (i) $F_\mathrm{sub}(\lambda_j)$ peaks significantly above the continuum; and (ii) $\xi_\mathrm{sub}$ calculated specifically at $\lambda_j$ (i.e., without binning) deviates significantly from the surroundings. The line candidates are cross matched to a galaxy emission line library to determine the precise redshift and eliminate false positives resulting from CRs and/or sky lines which are imperfectly removed.

Once the locations of prominent galaxy lines are determined, the spectral bins are assigned following a few strategies: (i) the maximum bin size is specified by the user (or 64 by default); (ii) the minimum bin size is limited by the spectral resolution $\Delta\lambda$; (iii) the finest bins are centered on each galaxy emission line, with bin sizes increasing progressively farther from the lines; and (iv) the widths of adjacent bins cannot differ by more than a factor of two. {An example of the adaptive binning is shown in the middle panel of Figure~\ref{fig:input}.}

\subsection{Building Priors from Archival Images}\label{sec:pipeline_prior}

\begin{figure*}
    \centering
    \includegraphics[width=\linewidth]{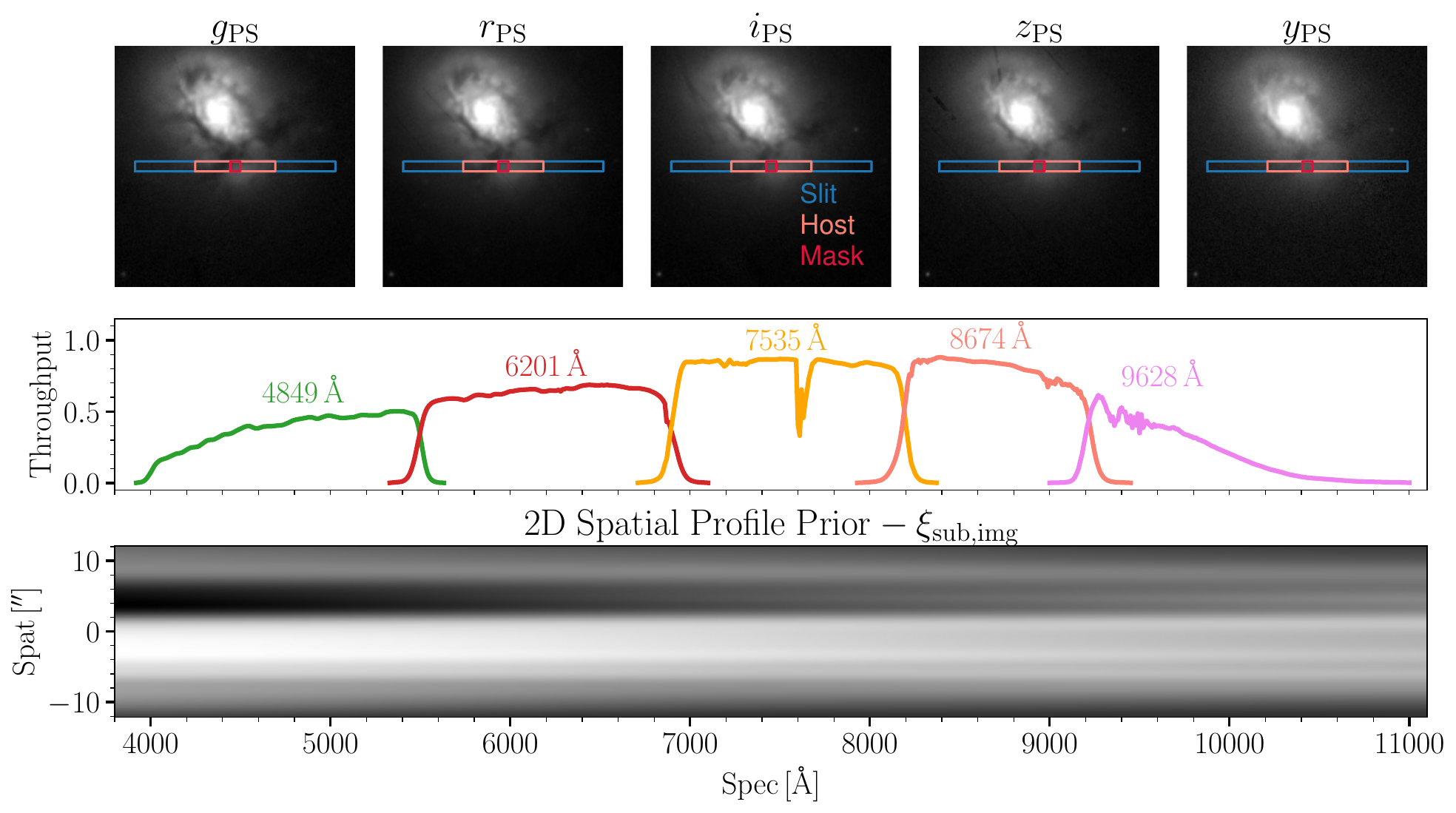}
    \caption{An example of constructing $\xi_\mathrm{sub,img}$, which we use as the mean function in modeling $\xi_\mathrm{sub}$ with a GP (Eq.~\ref{eq:2D_GP}). \textit{Top panels:} example archival PS images in $grizy$ filters. The position of a 1\arcmin-long slit is displayed as the blue patch. A 24\arcsec-wide host region and a transient mask with a width of twice the seeing are indicated as the salmon and red patches, respectively. 
    {\textit{Middle panel:} the throughout function of each filter and the corresponding effective wavelength.}
    \textit{Bottom panel:} the smooth $\xi_\mathrm{sub,img}$ built with the galaxy light profiles (within the host region) at five discrete wavelengths with a 2D GP.}
    \label{fig:host_prior}
\end{figure*}

Wide-field imaging surveys have visited almost every corner in the sky with multiple filters. \hostsub\ uses these images to build a smooth model of the host galaxy light distribution along the slit, which serves as the prior in modeling the 2D spectrum.

\hostsub\ automatically downloads image cutouts covering the transient and its host galaxy from a couple of different surveys, including PS, LS, and the shallower SDSS. Users are also encouraged to load their own higher-quality images. Images are processed via Astrometry.net \citep{Astrometry.net_2010} for precise astrometric calibration. Then they are rotated and resampled onto a grid centered at the location of the transient and oriented along the same direction of the slit when the spectrum is taken. The resampling is conducted using the \texttt{reproject\_adaptive} function of the \texttt{reproject} package \citep{reproject_2024}, which carries out anti-aliased reprojection {with} the \citet{DeForest_2004} algorithm. Using pixels covered by the slit of a specific width, \hostsub\ derives the mock spatial distribution of the galaxy flux along the slit. 
And by subtracting the mean flux in the same sky region defined in preprocessing the 2D spectrum (Eq.~\ref{eq:xi_sub}) and normalizing the flux residual (Eq.~\ref{eq:xi_sub_norm}), \hostsub\ obtains values of $\xi_\mathrm{sub,img}$ at a few grid points $(x_i, \lambda_j)$
{, effectively a transmission-weighted average of the spatial profile $\xi_\mathrm{sub}$, serving as an approximation to $\xi_\mathrm{sub}$ over the wavelength range covered by the filters.}
Notably, the spatial coordinates $x_i$ depend on the pixel scales in the archival images, 
{and we adopt the effective wavelength of each filter as a representative $\lambda_j$}. As there are at most a handful of broad-band filters available, the grid is expected to be coarse.

To build a smooth galaxy profile model out of the coarse grid, we leverage the 2D GP by assuming
\begin{equation}\label{eq:img_GP}
    \xi_\mathrm{sub,img}\sim\mathcal{GP}(\mu_\mathrm{img}, K_\mathrm{img}((x,\lambda),(x^\ast,\lambda^\ast))),
\end{equation}
where $\mu_\mathrm{img}$ is an unknown constant mean function, and $K_\mathrm{img}$ is a 2D kernel parameterized as the combination of Mat\'ern--3/2 kernel ($K_\mathrm{M32}$) and exponential squared kernel ($K_\mathrm{ES}$). With the same length scale $l$, the Mat\'ern-3/2 covariance declines faster than the exponential squared covariance, so qualitatively the Mat\'ern-3/2 kernel is better at capturing smaller scale fluctuations, whereas the exponential squared kernel better models any smooth, long term trend. The composite kernel is defined as\footnote{{In each bracket, the parameters before the semicolon denote the coordinates, while those after the semicolon denote the kernel hyperparameters.}}
\begin{equation}
    \begin{aligned}
        K_\mathrm{img} = &(A_{x,\mathrm{r}} K_\mathrm{M32}(x, x^\ast; l_{x, \mathrm{r}}) + A_{x,\mathrm{s}}K_\mathrm{SE}(x, x^\ast; l_{x, \mathrm{s}}))\\
        & \times (A_{\lambda,\mathrm{r}} K_\mathrm{M32}(\lambda, \lambda^\ast; l_{\lambda, \mathrm{r}}) + A_{\lambda,\mathrm{s}} K_\mathrm{SE}(\lambda, \lambda^\ast; l_{\lambda, \mathrm{s}})),
    \end{aligned}
\end{equation}
where $l_{x,\mathrm{r}}$ and $l_{x,\mathrm{s}}$ are the length scales on the spatial axis, $l_{\lambda,\mathrm{r}}$ and $l_{\lambda,\mathrm{s}}$ are the length scales on the spectral axis, while $A_{x,\mathrm{r}}$, $A_{x,\mathrm{s}}$, $A_{\lambda,\mathrm{r}}$, and $A_{\lambda,\mathrm{s}}$ parameterize the amplitudes and relative strength of the kernels.\footnote{{For simplicity, we omit the subscript ``img'' from all kernel parameters and adopt the same convention for the kernels defined in Eqs.~(\ref{eq:1D_GP}) and (\ref{eq:2D_GP}); note that these parameters are independent of one another.}}
Typically, we expect $l_{x,\mathrm{r}}\ll l_{x,\mathrm{s}}$ and $l_{\lambda,\mathrm{r}}\ll l_{\lambda,\mathrm{s}}$, so both the {rapid (r) and slow (s)} fluctuations on both the spatial and spectral directions can be well monitored. 

\hostsub\ allows users to impose reasonable constraints on the GP hyperparameters, particularly the length scale. {While the initial values and allowed ranges of all these kernel parameters could be specified in the configuration file, by default, we initialize $l_{\lambda,\mathrm{r}}$ and $l_{\lambda,\mathrm{s}}$ to be $10^4$\,\r{A} and $10^5$\,\r{A}, respectively, both with a lower bound of $10^3$\,\r{A} (which could be adjusted to match the wavelength range of the broad-band filters). Both $l_{x,\mathrm{r}}$ and $l_{x,\mathrm{s}}$ are initialized to be the typical seeing of the archival images, as} any real flux fluctuation along the spatial direction with a length scale shorter than the spatial resolution $\Delta x$, determined by the PSF, would have been smeared out and could not be physically meaningful. In ground-based observations, the PSF is primarily determined by atmospheric seeing. By default, \hostsub\ enforces a lower bound on the kernel length scale, requiring $l_{x,\mathrm{r}}, l_{x,\mathrm{s}}>\Delta x/2.355${, and an additional upper bound on $l_{x,\mathrm{r}}$ of $2\Delta x$}. The factor of 2.355 converts the full width at half maximum (FWHM) of the PSF, the typical astronomical measure of seeing, into the standard deviation of a Gaussian profile. This constraint effectively prevents the model from interpreting random noise as high-frequency spatial fluctuations. Figure~\ref{fig:host_prior} shows an example of constructing $\xi_\mathrm{sub,img}$ from archival PS images.

\hostsub\ implements the GP framework of \texttt{tinygp} \citep{tinygp_2024}, a lightweight and flexible GP library built on \texttt{JAX} \citep{jax_2018}, which accelerates numerical computations on Python arrays through just-in-time compilation, automatic differentiation, and GPU acceleration. The model hyperparameters are inferred by maximizing the marginal log-likelihood, corresponding to the log-probability of the GP conditioned on the discrete measurements of $\xi_\mathrm{sub,img}(x_i, \lambda_j)$, with gradient descent powered by \texttt{JAXopt} \citep{jaxopt_implicit_diff}. 

\subsubsection{Trace Alignment and 2D Co-addition}\label{sec:pipeline_co-add}
For faint transients embedded within bright host galaxies, direct object tracing in the original 2D spectrum is often unreliable. To address this, {\hostsub\ may also take} the trace of a bright, isolated reference source (e.g., a standard star), placed at the same location within the slit and observed with the same instrumental configuration, as an initial estimate of the trace. However, residual misalignment up to a few pixels can remain. {\hostsub} therefore {refines} the trace position by aligning the 2D spectrum to the spatial profile of the galaxy as modeled from archival imaging.
{On the raw 2D spectrum, the spatial profile is calculated as the flux-weighted, sigma-clipped average of the observed flux across a series of spatial bins in the host region. The pipeline then evaluates the imaging prior $\xi_{\mathrm{sub,img}}(x-x_{\mathrm{off}},\bar\lambda)$ at the flux-weighted mean wavelength $\bar\lambda$ of the 2D spectrum and searches a fine grid of trial offsets $x_{\mathrm{off}}$ for the value that maximizes the Pearson correlation between the two profiles. The resulting $x_{\mathrm{off,opt}}$ is applied as the trace correction, yielding sub-pixel alignment which substantially improves subsequent host modeling.}

The pipeline provides an option to co-add multiple rectified 2D spectra from individual exposures, as requested by users, to enhance signal-to-noise ratio. Prior to the co-addition, the median value of each row is subtracted from every 2D spectrum as a preliminary global sky subtraction, which helps to minimize background variations between exposures. {The pixel-wise CR rejection is also refined by contrasting multiple frames.}
The co-addition is performed as an inverse-variance weighted mean at each pixel, using only valid, unmasked data. The propagated error is computed accordingly from the weighted variances. 

\subsubsection{Seeing Matching}\label{sec:pipeline_seeing}
To match the 2D spectrum and the galaxy profile prior even more precisely, we also need to ensure that the spatial resolutions, determined by the seeing during the observations, are comparable. This is similar to the standard image subtraction, where the PSFs of the (new) science image and the reference image need to be precisely matched, typically by convolving an appropriate kernel (with an FWHM of $\Delta x_\mathrm{kernel}$) to degrade the spatial resolution of the sharper image \citep[e.g.,][]{Alard_1998},
\begin{equation}
    \Delta x_\mathrm{final}^2 = \Delta x_\mathrm{kernel}^2 + \Delta x_\mathrm{init}^2.
\end{equation}
Here $\Delta x_\mathrm{init}$ and $\Delta x_\mathrm{final}$ denote the initial and final seeing of the images, respectively.
Unlike imaging data, a 2D spectrum contains only one spatial dimension which can be used to decode the shape of the PSF. The PSF size also has a substantial negative wavelength dependence \citep{Boyd_1978}. Recovering the exact shape of the PSF in a 2D spectrum is particularly difficult. In this section, we show how the pipeline empirically matches the seeing of the 2D spectrum to that of the priors built with the archival images by minimizing the difference between the spectrum and the image.

If the seeing in the archival images is worse than in the spectroscopic data, the seeing matching is achieved by convolving the observed spectrum with 1D Gaussian kernels with a range of widths corresponding to possible seeing differences. For each case, the length scale of the kernel has a negative dependence on the wavelength, which can be approximated as a power law, $\sigma_\mathrm{kernel}\propto\lambda^{-\alpha}$.\footnote{We expect $\alpha=0.2$ if the atmospheric blurring is dominated by Kolmogorov turbulence \citep{Fried_1965, Woolf_1982}, but steeper dependencies can arise due to deviations from the Kolmogorov theory (e.g., local turbulence near the dome).}
The host profile in the convolved spectrum is then compared to the imaging prior, yielding the $\chi^2$ statistic, equivalent to the negative Gaussian log-likelihood. The optimal $\sigma_\mathrm{kernel}$ and $\alpha$ are inferred by minimizing the $\chi^2$, and then applied to the 2D spectrum. The allowed range of $\alpha$ is between 0.2 and 0.5. {After the matching process, $\xi_\mathrm{sub}$ is re-evaluated on the convolved spectrum, and the pipeline proceeds to the main GP modeling.}

Conversely, if the archival images have higher spatial resolution than the spectroscopic data, we match the seeing by degrading the archival images with 2D Gaussian kernels, again parameterized by $\sigma_\mathrm{kernel}$ and $\alpha$. For a ($\sigma_\mathrm{kernel},\alpha$) pair, kernel widths are computed for each image based on their effective wavelength, and the images are convolved with 2D Gaussian kernels of the corresponding widths. The pipeline then reconstructs the imaging model as described in Section~\ref{sec:pipeline_prior}, and similarly, the optimal $\sigma_\mathrm{kernel}$ and $\alpha$ are determined by minimizing the $\chi^2$. 

We note that our seeing matching process is likely an oversimplification. For example, since the spectroscopic and imaging data are usually collected at different observatories, their wavelength dependencies of seeing (i.e, $\alpha$) may differ. The goal of the seeing matching is not to build a perfect prior that exactly reproduces the spatial profile of the 2D spectrum, but rather to make the prior sufficiently similar at a reasonable computational cost. Our GP framework, illustrated in Section~\ref{sec:pipeline_main}, is then well suited for modeling the residuals, which are usually highly correlated.

\subsection{Main Process: GP Modeling}\label{sec:pipeline_main}
In the main module of the pipeline, we model $F_\mathrm{sub}$ and $\xi_\mathrm{sub}$ in a sequential way. In the first stage, \hostsub\ builds a preliminary model for $F_\mathrm{sub}$ with a 1D GP following Eq.~(\ref{eq:1D_GP}). {Between a pair of wavelengths $(\lambda, \lambda^\ast)$, the} kernel is defined as a combination of a Mat\'ern--3/2 kernel and a Mat\'ern--5/2 kernel, 
\begin{equation}\label{eq:kernel_1D}
    K_1 = A_\mathrm{r}K_\mathrm{M32}(\lambda,\lambda^\ast; l_{\lambda, \mathrm{r}}) + A_\mathrm{s}K_\mathrm{M52}(\lambda,\lambda^\ast; l_{\lambda, \mathrm{s}}),
\end{equation}
which captures both small-scale fluctuations near narrow nebular emission and absorption lines, as well as large-scale trends across the continuum. To enable scalable GP inference, we adopt quasi-separable kernels implemented in \texttt{tinygp},\footnote{The exponential squared kernel is not quasi-separable, so we use the Mat\'ern--5/2 kernel implementation in the \texttt{tinygp.quasisep} package, which is smoother than the Mat\'ern--3/2 kernel, to characterize large-scale variations.} reducing the computational complexity from $\mathcal{O}(N^3)$ to $\mathcal{O}(N)$ on 1D datasets \citep{Eidelman1999, Foreman-Mackey_2017}. By default, both $l_{\lambda, \mathrm{r}}$ and $l_{\lambda, \mathrm{s}}$ are forced to be greater than the spectral resolution $\Delta \lambda/2.355$. For the {more rapidly} varying kernel, an upper bound of $2\Delta \lambda$ is imposed on $l_{\lambda, \mathrm{r}}$. The point estimates of the 1D GP hyperparameters are then obtained by maximizing the {marginal} likelihood.

Once a reasonable model for $F_\mathrm{sub}$ is obtained, the pipeline proceeds to a second stage where both
the 1D $F_\mathrm{sub}$ and the 2D $\xi_\mathrm{sub}$ are modeled jointly. The optimal hyperparameters derived in the first stage are adopted as initial values for this round. The mean function $\xi_\mathrm{sub,img}$ in Eq.~\ref{eq:2D_GP} has been described in detail in Section~\ref{sec:pipeline_prior}. The covariance matrix is constructed using a 2D Mat\'ern--3/2 kernel
\begin{equation}\label{eq:kernel_2D}
    K_2 = A_\mathrm{M32}K_\mathrm{M32}((x, \lambda),(x^\ast, \lambda^\ast); l_{x}, l_{\lambda}).
\end{equation}
As in the previous GP modeling step, \hostsub\ imposes lower bounds on the kernel length scales, requiring $l_x > \Delta x/2.355$ {to} prevent the model from fitting noise fluctuations that occur below the spatial resolution limits. {Similar as that in defining the prior kernel in Section~\ref{sec:pipeline_prior}, by default, we set the lower bound of $l_\lambda$ to be $10^3$\,\r{A}.}

As discussed in Sections~\ref{sec:pipeline_host_lines}, $\xi_\mathrm{sub,img}$ is typically not an adequate approximation of $\xi_\mathrm{sub}$ in the vicinity the nebular emission lines from the host galaxy, where abrupt jumps in $\xi_\mathrm{sub}$ can occur along the spectral axis {(see also Figure~\ref{fig:input})}. In such regions, the covariance structure of $\xi_\mathrm{sub}$ should account for the local discontinuities, with reduced correlation across wavelengths that straddle an emission line feature.
Conversely, when both pixels lie within the vicinity of the same emission line, the covariance between their corresponding values of $\xi_\mathrm{sub}$ may be enhanced. This reflects the empirical observation that the spatial structure of the galaxy profile is often more variable near nebular emission lines, owing to the clumpy and irregular distribution of star-forming regions compared to the smoother profile associated with older stellar populations.
To account for these discontinuities, we introduce a custom multiplicative kernel term $k_\mathrm{corr}$. For each emission line centered at $\lambda_{\mathrm l_i}$, we define a proximity function using a hyperbolic tangent window,
\begin{equation}
\begin{aligned}
    w_{\mathrm l_i}(\lambda;, l_\lambda) = \frac{1}{2} \Bigg[ & \tanh\left(\frac{2(\lambda - \lambda_{\mathrm l_i} + l_\lambda/2)}{l_\lambda}\right) \\
    - & \tanh\left(\frac{2(\lambda - \lambda_{\mathrm l_i} - l_\lambda/2)}{l_\lambda}\right) \Bigg],
\end{aligned}
\end{equation}
which serves as a differentiable approximation to the indicator function $\mathbf{1}[|\lambda - \lambda_{\mathrm l_i}| < l_\lambda/2]$. This formulation enables the use of automatic differentiation for optimizing the width parameter $l_\lambda$, which sets the effective width influenced by the emission line. In the optimization, the width parameter is also required to be greater than $\Delta\lambda$. We then define two auxiliary functions $W_\mathrm{one}$ and $W_\mathrm{both}$:
\begin{equation}
\begin{aligned}
    W_{\mathrm{one}, \mathrm l_i} =\ 
    & w_{\mathrm l_i}(\lambda; l_\lambda) \cdot \left[1 - w_{\mathrm l_i}(\lambda^\ast; l_\lambda)\right] \\
    +\ & \left[1 - w_{\mathrm l_i}(\lambda; l_\lambda)\right] \cdot w_{\mathrm l_i}(\lambda^\ast; l_\lambda)
\end{aligned}
\end{equation}
which attains values close to 1 if and only if exactly one of the two points lies near the emission line; and
\begin{equation}
\begin{aligned}
    W_{\mathrm{both}, \mathrm l_i} =\ 
    w_{\mathrm l_i}(\lambda; l_\lambda) \cdot w(\lambda^\ast; l_\lambda)
\end{aligned}
\end{equation}
which approaches 1 only when both points are within the vicinity of the same line. The emission line correction factor is defined as
\begin{equation}
\begin{aligned}
    k_{\mathrm{corr}} = 
    \prod_i \Big[ 
    & 1 - W_{\mathrm{one}, \mathrm l_i}(\lambda, \lambda^\ast; l_\lambda) \\
    & + A_\mathrm{l} \cdot W_{\mathrm{both}, \mathrm l_i}(\lambda, \lambda^\ast; l_\lambda) 
    \Big],
\end{aligned}
\end{equation}
where $A_\mathrm{l} > 0$ controls the amplitude of the enhancement when both points are near the same emission feature. For all emission lines, $A_\mathrm{l}$ is assumed to be the same. Eventually, the 2D covariance kernel is corrected as
\begin{equation}
    K_2' = k_{\mathrm{corr}}(\lambda, \lambda^\ast; A_\mathrm{l}, l_\lambda)\cdot K_2.
\end{equation}
{The correction term $k_\mathrm{corr}$, as well as the adaptive batching strategy, effectively isolates the modeling of $\xi_\mathrm{sub}$ inside and outside the emission line regions, and helps to prevent the GP from being influenced by discontinuities at emission lines when modeling the smoother continuum regions.}

The total log-likelihood to be maximized consists of the marginal log-likelihood functions from the GPs, along with a Gaussian log-likelihood term based on the residuals between the modeled and observed flux, $\widehat f_\mathrm{sub}$ and $f_\mathrm{sub}$, summed over all pixels
\begin{equation}
    \ln \mathcal{L} = \ln \mathcal{L}_{\mathcal{GP}, 1} + \ln \mathcal{L}_{\mathcal{GP}, 2} - \sum_{x_i, \lambda_j}\frac{(\widehat f_\mathrm{sub} - f_\mathrm{sub})^2}{2\sigma_\mathrm{sub}^2},
\end{equation}
where {$\mathcal{L}_{\mathcal{GP}, 1}$ and $\mathcal{L}_{\mathcal{GP}, 2}$ correspond to the marginal likelihood functions given the observed $F_\mathrm{sub}$ and $\xi_\mathrm{sub}$ under the GP priors \citep{Aigrain_GP_2023}, as characterized by Eqs.~(\ref{eq:1D_GP}) and (\ref{eq:2D_GP}), respectively, and} $\sigma_\mathrm{sub}$ represents the per-pixel uncertainty in $f_\mathrm{sub}$, assuming independent Gaussian noise. Since the log-likelihood is maximized under parameter constraints (e.g., lower bounds on kernel length scales and any other customized constraints imposed by the users), the resulting point estimates of {kernel} hyperparameters correspond to maximum a posteriori estimators under implicit priors.

Now that both the global and local sky background $\langle f_\mathrm{obs}\rangle_\mathrm{sky}$ and $f_\mathrm{sub}$ have been derived for the rectified 2D spectrum, the ultimate sky model $\langle f_\mathrm{obs}\rangle_\mathrm{sky}+f_\mathrm{sub}$ is evaluated on each pixel on the original 2D spectrum as the final output of the pipeline. Downstream tasks, including 1D spectrum extraction and fluxing, may then be performed with external pipelines.

\section{Results}\label{sec:result}
\subsection{Test on a Synthetic Dataset}\label{sec:result_syn}
\begin{figure*}
    \centering
    \includegraphics[width=\linewidth]{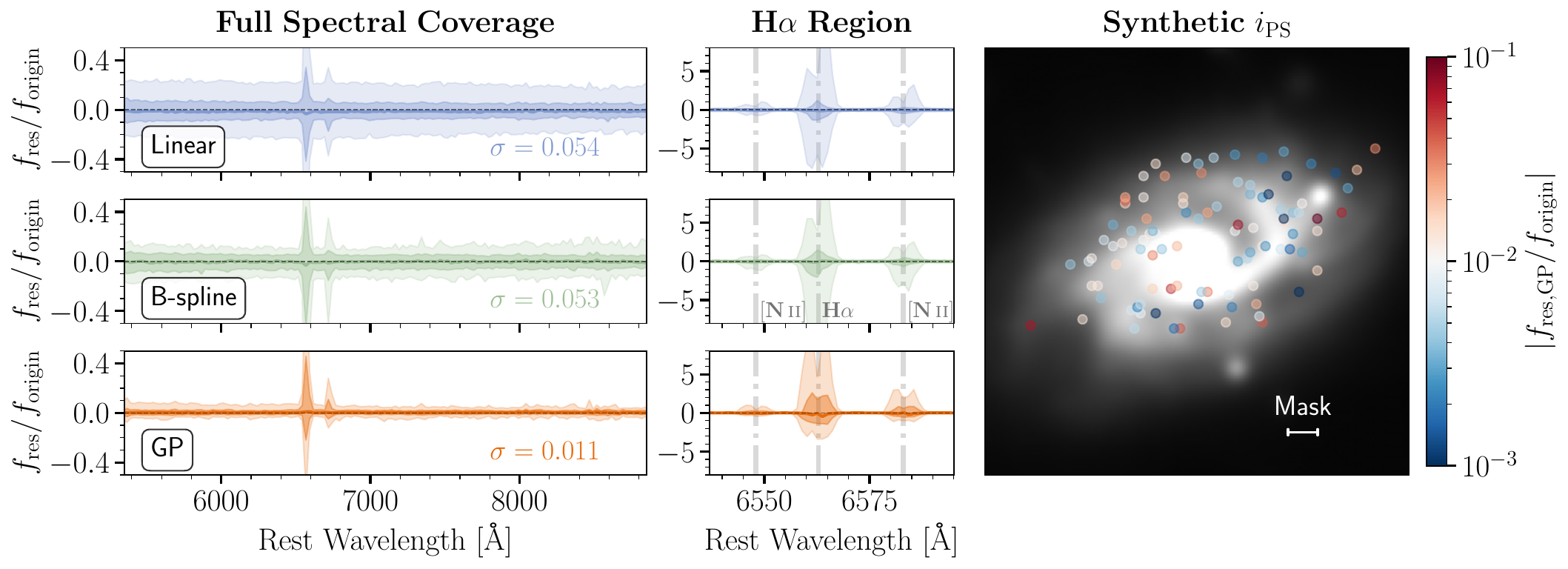}
    \caption{The GP method systematically improves the galaxy background estimates on the MUSE synthetic dataset of the spiral galaxy LCRS B110709.2-121854. 
    {The first two columns illustrate the distribution of fractional flux residuals (residuals normalized by the original galaxy flux in the PS $i$ filters at randomly drawn slit locations) as a function of wavelengths in each of the 25\,\r{A} bins, resulting from the classic linear and B-spline interpolation methods, and the GP method. The first column covers the full wavelength range, whereas the second column zooms in on the region containing the strong H$\alpha$ and [\ion{N}{2}] emission lines. Solid lines denote the median residuals across 100 random draws; the darker and lighter shaded regions represent the 68\% and 95\% percentiles, respectively. As a comparison, the black dashed line denotes the zero residual level.
    The $\sigma$ values in the legends quantify the standard deviation of the 100 median fractional residuals, each computed across the full wavelength range for a single random draw.}
    The rightmost panel shows the synthetic image in the $i$ band convolved with a Gaussian kernel to downgrade the spatial resolution {from 0.68\arcsec} to 1.2\arcsec. Overlaid are the locations of the randomly drawn slits, color coded by the corresponding absolute values of fractional residuals {(median across all wavelengths)} from the GP method. The typical aperture mask width (twice the seeing, {2.4\arcsec}) is indicated, though the mask width is determined case by case in the seeing matching.
    }
    \label{fig:muse_spiral}
\end{figure*}
\begin{figure*}
    \centering
    \includegraphics[width=\linewidth]{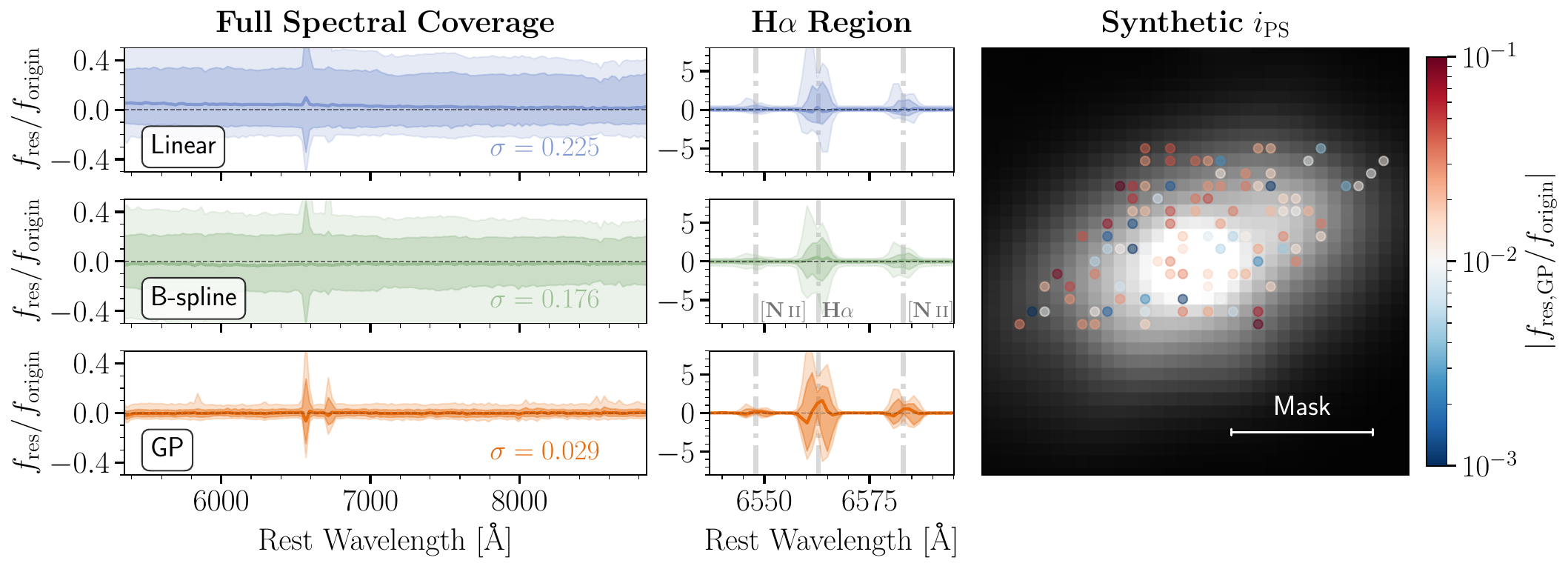}
    \caption{
        Same as Figure~\ref{fig:muse_spiral}, but showing that the GP method still outperforms the classic method{s in both the precision and accuracy} on the synthetic dataset where the MUSE hyper-spectral data cube has been spatially binned by a factor of 4{, simulating observations of a less resolved galaxy scaled to 25\% of the angular size under the same seeing conditions.} Again, seeings in the synthetic photometry on the rightmost panels have been downgraded to $\sim$1.2\arcsec.
    }
    \label{fig:muse_spiral_bin}
\end{figure*}
To evaluate the effectiveness of this new technique in improving host galaxy subtraction relative to the classic aperture spectroscopic method, we validate our pipeline using a large set of 2D synthetic galaxy spectra. Since these spectra do not include any transient, an ideal host subtraction method should produce zero flux residuals at all wavelengths. The synthetic 2D spectra are generated from a 3D hyper-spectral data cube acquired with the Multi Unit Spectroscopic Explorer \citep[MUSE;][]{MUSE_2010}, which encodes the galaxy spectrum at every pixel within the field of view. By placing an artificial ``slit'' on the data cube, we could construct a 2D spectrum by aligning the 1D spectra from pixels on the slit.

We test our package on the archival MUSE data of LCRS B110709.2-121854 (PI: Tanvir), a galaxy containing a bright, red bulge (FWHM $\sim$ 2.5\arcsec) and well resolved spiral arms. We randomly place an artificial slit at 100 locations, with a width of 1\arcsec. For simplicity, we only draw slits along the columns of the data cubes. The 2D spectrum is the column-average of the 1D spectrum for each row. Then a mask with an aperture size of twice the seeing placed on the 2D spectrum, where the mask center is also drawn randomly. Surrounding the mask is the host region to be modeled which is $10\arcsec$ wide. Note that the MUSE field is slightly distorted, and hence the alignment with the real archival images is non-trivial. We therefore build our priors using the synthetic photometry in the PS $r$, $i$, and $z$ filters by convolving the 3D data cube with the response curves of PS. The MUSE observations were conducted with a seeing of {$\Delta x=0.68\arcsec$}, much better than the typical seeing in PS1. To mimic the common situation where the seeing in the archival images is worse than that in the spectroscopic observations, we convolve a wavelength-dependent Gaussian kernel with the synthetic images. The FWHM of the Gaussian kernel is {$\Delta x_\mathrm{kernel}=1\arcsec$}\ in the PS $r$ band, and the corresponding downgraded seeing is $\sim$1.2\arcsec.  Adopting a wavelength dependence of $\lambda^{-0.2}$, the FWHMs of the kernels applied to the $i$ and $z$ bands are 0.96\arcsec\ and 0.93\arcsec, respectively.

In Figure~\ref{fig:muse_spiral} we show the distribution of the median galaxy flux residuals {as a function of wavelength in each of the 25\,\r{A} bins}. The residual ($f_\mathrm{res}$) is normalized by the galaxy background {in the synthetic PS $i$ band} ($f_\mathrm{origin})$, and should be considered as the fractional flux residual. 
{Across the entire wavelength range}, the GP method (orange) significantly improves the subtraction {of the host galaxy continuum} compared to the two classic methods, which estimates the galaxy background {by interpolating} the flux in the local sky regions {surrounding the mask, either linearly (blue) or with a B-spline (green). In the linear-interpolation case, each part of the local sky region spans 1--2\,$\Delta x$ from the trace center. In the B-spline approach, we fit a third-order B-spline to a broader region, each part spanning 1--6\,$\Delta x$ from the trace center (or up to the slit edge), and set knot intervals to be $2\Delta x$.} 
Statistically, the GP method reduces the scatter of the fractional residuals by a factor of $\sim${5, and only 5\% of the experiments have fractional residuals larger than 5\%}, 
all without custom hyperparameter tuning. {In contrast, with the classic linear/B-spline subtraction method, 45\%/37\% of the experiments yield fractional residuals exceeding 5\%.} 
{On the other hand, the GP method still cannot remove narrow galaxy emission lines (second column), but the strong over-/under-subtraction is limited to the immediate vicinity of the lines and does not affect the continuum modeling.} 
The rightmost panel displays the locations of the slit centers, color coded with the corresponding absolute values of fractional flux residuals from the GP method. {The fractional flux residuals exhibit no significant spatial pattern, whereas the absolute flux residuals are elevated near the galaxy center with the highest background.}

We further assess the model performance when the galaxy is less resolved. In practice, we reduce the spatial resolution by performing 4$\times$4 binning to the data cube, but retain the original pixel scale of 0.2\arcsec/pixel. {Again, we convolve Gaussian kernels with the synthetic $riz$ images to worsen the seeings to $\sim$1.2\arcsec.} The results are displayed in Figure~\ref{fig:muse_spiral_bin}. Effectively, in {this test} the galaxy spans only 25\% of its original angular size. {The} bulge and spiral arms are no longer resolved and the FWHM of the galaxy outskirt is $\sim$2\arcsec, only marginally resolved with the worsened seeing. We find that the scatter of the fractional residuals increases as the galaxy gets less resolved, but the GP method outperforms the classic method{s even more significantly in this case, reducing the scatter by a factor of $\sim$6--8. More notably, the linear method yields substantial biases in flux residuals, with the median values tending to be positive, indicating systematic under-subtraction of galaxy background due to the overall curvature of the galaxy profile. In comparison, the GP method effectively corrects for this bias.}
Eventually, the GP method hits the limit when the physical angular size of the galaxy is comparable with the seeing.

Following the success on the synthetic dataset, in the next sections we will apply our pipeline to two real transients observed with the Low-Resolution Imaging Spectrometer \citep[LRIS;][]{LRIS_1995} on Keck, in which host galaxy removal critically affects our science interpretations of the spectra.

\subsection{Revisiting a Nebular-phase SN: SN\,2019eix}\label{sec:result_19eix}
\begin{figure*}
    \centering
    \includegraphics[width=\linewidth]{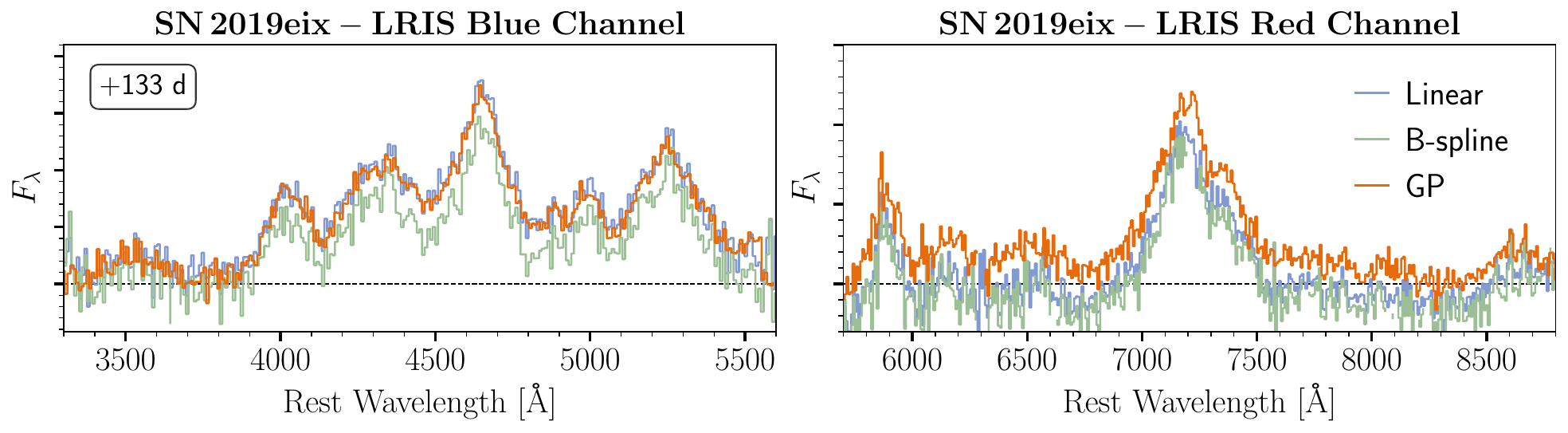}
    \caption{The GP method (orange) effectively removes the host galaxy contamination in the nebular-phase spectrum of SN\,2019eix, whereas the spectrum extracted with the classic methods (blue and green) {suffers an over-subtraction, which hinders detailed emission line modeling. All spectra have been smoothed and binned to  a 7.5\,\r{A} resolution. The horizontal dashed line indicates zero flux level. The flux scales are different between the two panels.} }
    \label{fig:19eix}
\end{figure*}

Hundreds of days after an SN explodes, the ejecta become optically thin. In this nebular phase, the spectra are dominated by the forbidden emission lines of the ions in the ejecta interior. Nebular-phase emission features uniquely probe the nucleosynthesis yields, ionization states, and kinematics in the innermost region of the SN ejecta, which are sensitive to the explosion channels and progenitor properties of both core-collapse and thermonuclear SNe.

However, SNe are naturally very faint in the nebular phase meaning they are often  outshone by host galaxy light even in low-contamination environments. Here we revisit SN\,2019eix, a peculiar SN with a vague classification \citep{Gonzalez_19eix_2023}, and show that a nebular-phase spectrum with precise host galaxy subtraction provides smoking-gun evidence of its explosion mechanism and progenitor nature.

SN\,2019eix is a hydrogen-poor SN that exploded near the bulge of NGC6695 ($z=0.0183$) on its bar. SN\,2019eix was originally classified as an SN\,Ic \citep{Dimitriadis_19eix_2019}, i.e., the explosion of a massive star with an envelope that has been stripped of hydrogen and helium. Yet \citet{Gonzalez_19eix_2023} demonstrate that the spectral sequence of SN\,2019eix are also consistent with some subluminous SNe\,Ia as outcomes of an exploding white dwarf (WD), based on a dataset up to $\sim$30\,days after maximum luminosity. As is shown in the Figure~7 in \citet{Gonzalez_19eix_2023}, around the maximum luminosity of SN\,2019eix, its spectral shape resembles both particular subluminous SNe\,Ia (e.g., SN\,1991bg; \citealp{Filippenko_91bg_1992}, SN\,2016hnk; \citealp{galbany_16hnk_2019, jacobson-galan_16hnk_2020}, and SN\,2018byg; \citealp{de_18byg_2019}), and SNe\,Ic \citep[e.g., PTF12gzk;][]{Ben-Ami_gzk_2012}. At this phase, differences between these peculiar events are subtle -- on average SNe\,Ic exhibit stronger \ion{Si}{2}/\ion{O}{1} ratios than SNe\,Ia \citep{Gal-Yam_2017, Sun_2017}, but the distribution of this diagnostic is a continuum and typing can be uncertain for objects near the boundary.

About 133\,days after maximum light, SN\,2019eix was observed by Keck/LRIS (PI: Fremling). We show the spectrum in Figure~\ref{fig:19eix}{, which} is dominated by broad emission features. {Notably, using classic interpolation-based methods\footnote{Pipelines such as \texttt{PypeIt} also employ B-splines to model the local sky, and the results can be sensitive to the specific implementation, including how the local sky region is defined for the fit and how many knots are used, so it is tricky to recreate the methodology in other pipelines. Throughout this paper, we use the same implementations of both the linear and B-spline methods described in Section~\ref{sec:result_syn} for all comparisons.} (linear and B-spline), the galaxy background in the LRIS red channel is overestimated, so a substantial fraction of the spectrum shows negative flux.}

\begin{figure*}
    \centering
    \includegraphics[width=\linewidth]{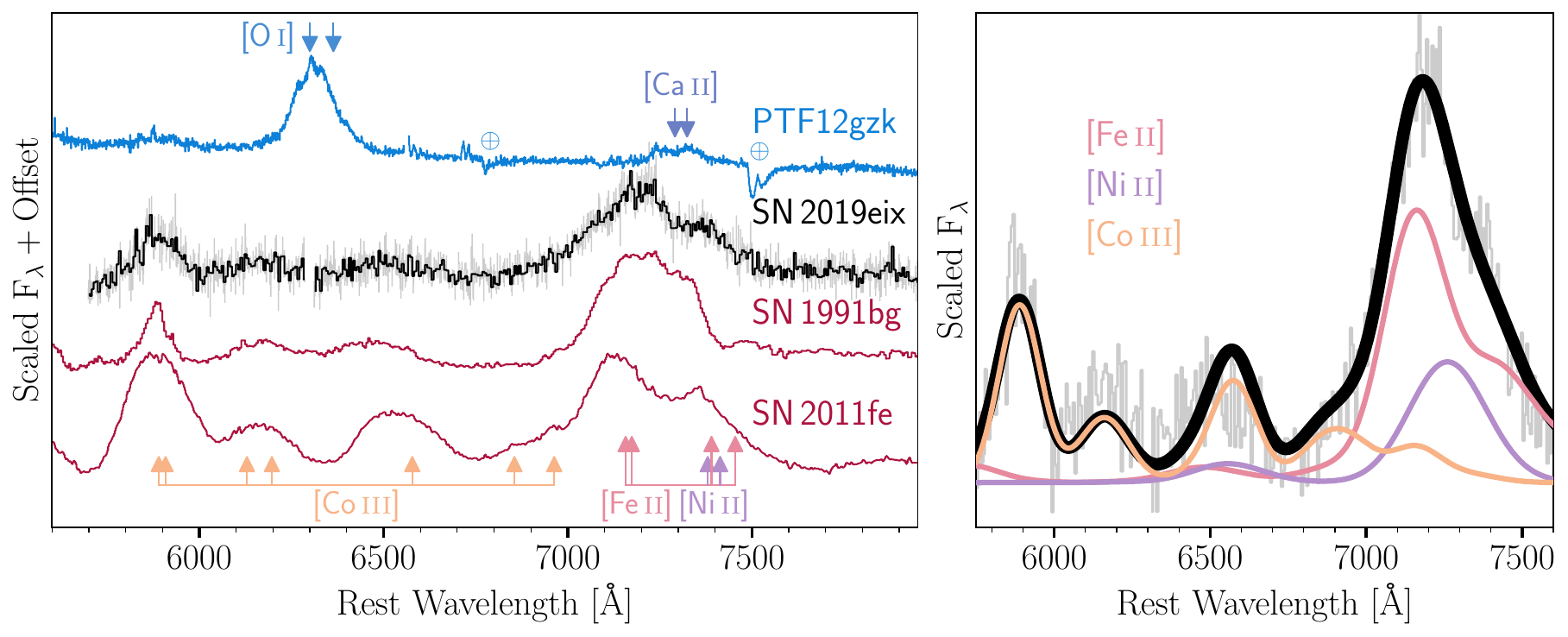}
    \caption{The nebular-phase spectrum of SN\,2019eix reveals its sub-{Chandrasekhar-mass} WD progenitor. \textit{Left:} the late-time spectrum of SN\,2019eix highly resembles those of the normal SN\,Ia 2011fe \citep{Stahl_2020} and the sub-luminous SN\,Ia 1991bg \citep{Filippenko_91bg_1992} (both in maroon). The gray spectrum is the raw SN\,2019eix spectrum with the host galaxy subtracted using the GP method, and the black spectrum is smoothed to the resolution of 5\,\r{A}. Within this wavelength range, spectra of SNe\,Ia are dominated by numerous [\ion{Fe}{2}], [\ion{Ni}{2}], [\ion{Co}{3}] lines, where the strongest lines are indicated. On the other hand, the spectrum of the SN\,Ic, PTF12gzk \citep{Ben-Ami_gzk_2012} (blue) only shows [\ion{O}{1}]\,$\lambda\lambda$6300, 6364 and [\ion{Ca}{2}]\,$\lambda\lambda$7291, 7324. The telluric features in the PTF12gzk are indicated with $\oplus$. \textit{Right:} modeling the 7300\,\r{A} complex of SN\,2019eix{, following the methodology in \citet{Liu_20jgb_2023, Liu_22joj_2023},} yields a low {[Ni\,{\sc ii}]/[Fe\,{\sc ii}] ratio ($\sim$0.5),} suggesting a sub-{Chandrasekhar-mass} WD origin. For a clean visualization we only display the smoothed spectrum in gray. The solid black curve is the joint model, whereas each colored curve corresponds to the contribution of each ion. 
    }
    \label{fig:19eix_comp}
\end{figure*}

In Figure~\ref{fig:19eix_comp} we show a comparison of SN\,2019eix with a few other SNe\,I in the nebular phase. Despite their similarity near peak to SNe\,Ic, the nebular-phase spectra of these SNe\,Ia serve as a clear discriminant between the two classes. 
{In both normal and subluminous SNe\,Ia, such as SN\,2011fe and SN\,1991bg, the nebular-phase spectra are dominated by forbidden emission lines of iron, nickel, and cobalt.}
On the contrary, the {spectra} of {hydrogen-poor} core-collapse {SNe, such as PTF12gzk, are} dominated by the [\ion{O}{1}] doublet.
As for SN\,2019eix, prominent iron, nickel, and cobalt emission lines are revealed in the nebular-phase spectrum ($+133$\,days) after careful host galaxy subtraction with the GP method, while there is no detectable [\ion{O}{1}], as expected in most SNe\,Ia. Additionally, the prominent line complex between 4000 and 5000\,\r{A} is another hallmark of SNe\,Ia. This emission complex, which is dominated by [\ion{Fe}{3}], is not nearly as strong in core-collapse SNe (Figure~\ref{fig:19eix}).  This confirms that SN\,2019eix originates from a thermonuclear explosion of a WD.

While the classic subtraction methods can already reveal the major emission lines, the over-subtraction of galaxy light hinders accurate line intensity measurements, and more precise host galaxy subtraction is required to place stringent constraints on the progenitor properties. From the early-time observables, SN\,2019eix has been suggested to be a peculiar SN\,Ia arising from the helium-shell double detonation of a sub-Chandrasekhar-mass WD \citep{Gonzalez_19eix_2023}, which is independently supported by the low [\ion{Ni}{2}]/[\ion{Fe}{2}] ratio in the 7300\,\r{A} line complex \citep{Maguire_2018, Flors_2020,Blondin_2022} from the host-subtracted nebular-phase spectrum presented here (Figure~\ref{fig:19eix_comp}). 

\subsection{Revisiting a Nuclear Transient: AT\,2019qiz}\label{sec:result_19qiz}
\begin{figure*}
    \centering
    \includegraphics[width=\linewidth]{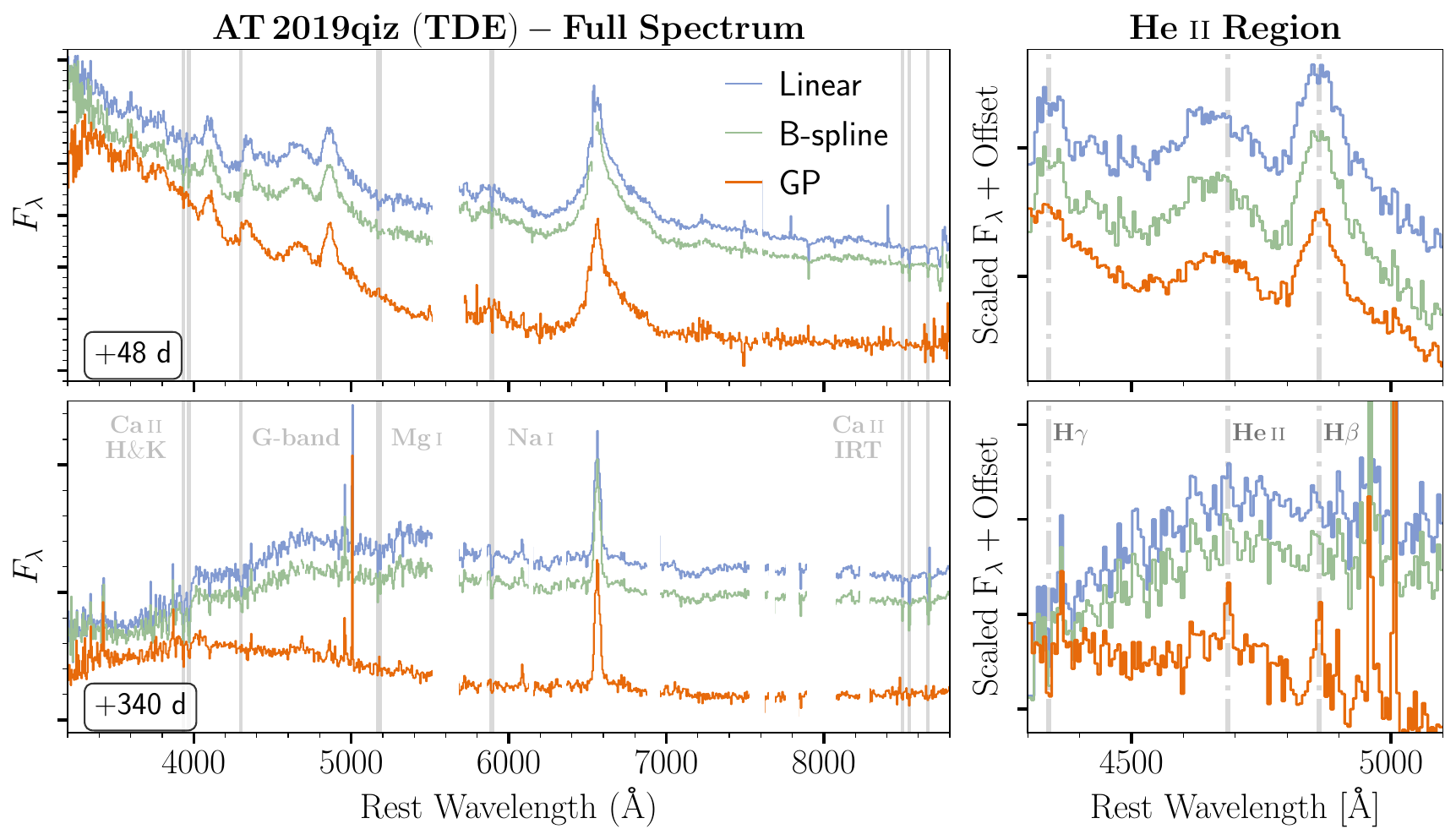}
    \caption{{The GP method effectively removes the host galaxy contamination for AT\,2019qiz,} a bright TDE at the nucleus of a nearby galaxy. {The left panel shows the full spectra (binned into 5\,\r{A}), with some common galaxy absorption lines -- the \ion{Ca}{2}\,H\&K doublet, G-band, the \ion{Na}{1}\,D doublet, the \ion{Mg}{1}~$\lambda\lambda\lambda$5167, 5173, 5184 triplet, and the Ca infrared triplet (IRT) -- labeled in gray, all of which are better removed with the GP method}. The right panel zooms in around the broad \ion{He}{2}\,$\lambda$4686 emission. With careful host galaxy subtraction using the GP method, we fix the underlying continuum beneath the \ion{He}{2}\,$\lambda$4686 and H$\beta$ emission in the 2019 spectrum ($+48$\,days), critical for estimating their relative intensity, and uniquely show that they still persist in the 2020 spectrum ($+340$\,days). We also probe the emerging narrow \ion{He}{2} and H$\beta$ emission (dash-dotted lines) in the 2020 spectrum that is not {obvious in the spectra reduced by the linear or B-spline method}. The same galaxy absorption lines from Figure~\ref{fig:19eix} are shown as vertical gray lines. Note that this spectrum consists of a single 900\,s exposure, so the red side is heavily contaminated by CRs; we therefore manually masked the affected pixels. Emission lines of [\ion{O}{3}] at 4363, 4959, and 5007\,\r{A} are detectable in the 2020 spectrum, but are not labeled for clarity. }
    \label{fig:19qiz}
\end{figure*}
\begin{figure*}
    \centering
    \includegraphics[width=\linewidth]{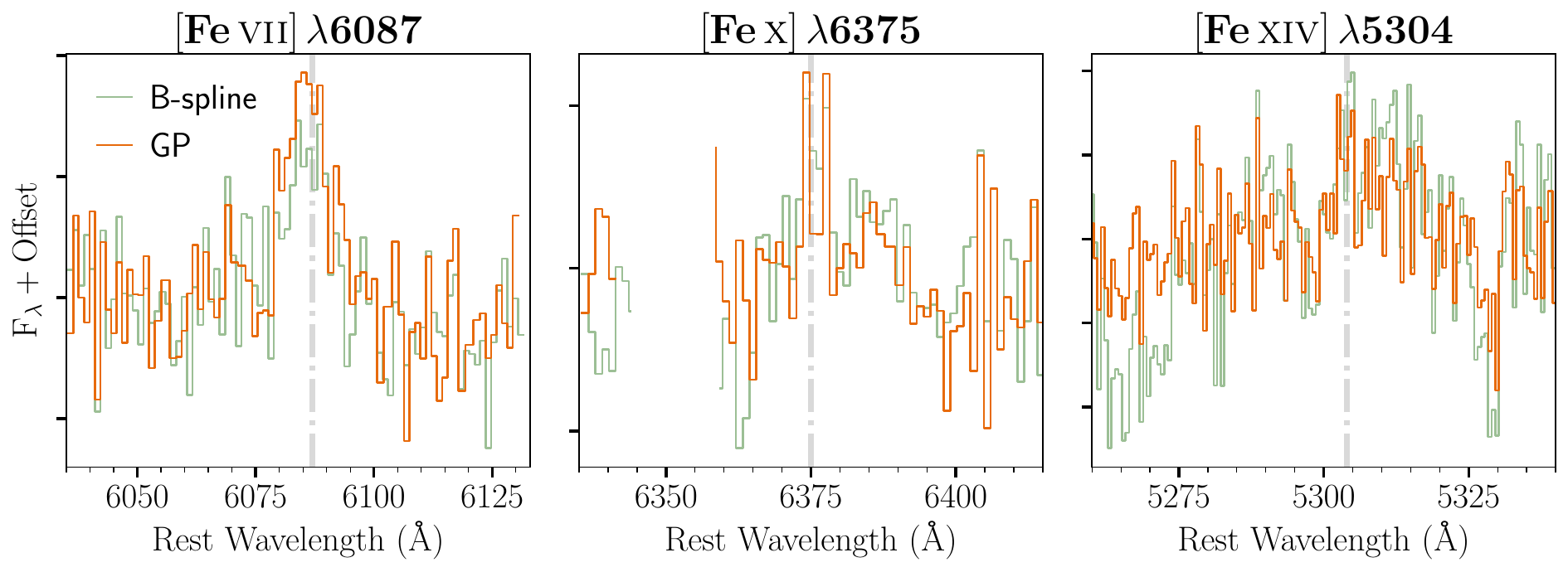}
    \caption{The GP method recovers smooth TDE continua near Fe coronal lines, as later identified in \citet{Short_2023}, from the 2020 LRIS spectrum of AT\,2019qiz. This enables robust flux measurements of the lines from the source. Spectra reduced with both the {classic B-spline} and the GP method are shown with an added offset for alignment, but fluxes are not rescaled. Central wavelengths of these coronal lines are labeled as the {dash-dotted} gray lines. The [\ion{Fe}{7}]\,$\lambda$6087 emission, isolated from strong galaxy features, appears consistent in both reductions. However, [\ion{Fe}{10}]\,$\lambda$6375 and [\ion{Fe}{14}]\,$\lambda$5304 lie near prominent galaxy absorption features, and only the GP method recovers the smooth continua. The region around [\ion{Fe}{11}]\,$\lambda$7892 is not displayed due to severe CR contamination.}
    \label{fig:19qiz_fe}
\end{figure*}
Tidal disruption events (TDEs) originate from the disintegration of stars grazing the tidal radius of a super-massive black hole (SMBH) in their strong tidal field. They are unique probes of both the demographics of the SMBH properties (e.g., mass) and the nuclear stellar population. TDEs spotted by all-sky time-domain surveys are UV-bright transients featured by a blue continuum and broad hydrogen, helium, and nitrogen emission lines in the optical spectra, though a puzzling subclass of TDEs appear to be featureless throughout their evolution. Yet as nuclear transients, host galaxy subtraction is extremely challenging for spectroscopic observations of TDEs. Traditionally, a spectrum containing light from the TDE and the galaxy nucleus is first extracted, and then the host contamination is subtracted in the post-processing by fitting the joint spectrum as a combination of the galaxy spectrum (a rescaled archival spectrum of the galaxy, if it exists, or a synthetic spectral model), and a TDE continuum model (blackbody or low-order polynomial). Empirical free parameters are introduced to handle the systematics due to the unknown slit loss in the long-slit spectroscopy, host galaxy extinction, etc.
Our technique provides an alternative, self-consistent way of removing the host contamination for TDEs in nearby galaxies. Here we revisit the spectroscopy of AT\,2019qiz to show how we are able to recover weak emission features of a TDE after careful host subtraction.

AT\,2019qiz is a H+He TDE with prominent broad hydrogen and helium emission lines \citep{Nicholl_2020, Hung_2021}. Its optical spectral sequence is featured by the delayed appearance of narrow nebular lines of H and \ion{He}{2}, as well as coronal lines of highly ionized Fe \citep{Short_2023}. Hosted by a nearby galaxy ($z=0.0151$) whose nucleus (FWHM $\sim$ 2\arcsec) can be marginally resolved by ground-based facilities, AT\,2019eix is an ideal object to test \hostsub. We re-reduced two Keck/LRIS spectra (PI: Foley) in \citet{Hung_2021} obtained 48\,days and 340\,days after the maximum luminosity of the TDE, both displayed in Figure~\ref{fig:19qiz}. The 2020 spectrum was obtained with a single, long (900\,s) exposure and is heavily contaminated by CRs, so in Figure~\ref{fig:19qiz} we have omitted the affected pixels.\footnote{Even if the trace of the transient is not directly impacted by CRs, if a significant portion of pixels in its vicinity is contaminated, the GP method will struggle to model the neighboring 2D spectrum. Therefore \hostsub\ is more sensitive to CRs than classic methods.}

{With} the GP method, we dramatically reduce the level of host contamination compared to {classic linear subtraction technique}. {In the 2019 spectrum, t}he Balmer break and the strong Ca H\&K absorption lines are {much better} removed, reshaping the continuum between $\sim$4000--5000\,\r{A}, where multiple broad H and He emission lines are present. This simplifies the measurement of the line flux, as an external galaxy spectral model is no longer required. In the 2020 spectrum, we uniquely detect (i) persistent broad \ion{He}{2}\,$\lambda$4686 emission; (ii) the emergence of narrow \ion{He}{2} and H$\beta$ emission -- notably, using the classic subtraction technique, the narrow \ion{He}{2} is not detected until the last epoch of the VLT/X-shooter spectrum from \citet{Short_2023}; (iii) the appearance of [\ion{Fe}{7}]\,$\lambda$6087, and unambiguous non-detections for other coronal lines of higher ionization states (see Figure~\ref{fig:19qiz_fe}), by removing prominent line features from the galaxy and recovering smooth continua around lines of interest. It has been shown that these recombination lines keep brightening over time \citep{Short_2023}, as the ionizing photons from the initial outburst reach material at larger radii. The fluxes of these recombination lines uniquely probe the gas properties surrounding the SMBH as well as the SED of the TDE outburst in the EUV and soft X-ray ($\sim$0.05--0.5\,keV) range. With the GP method, we achieve more secure detections of these key recombination lines, enabling robust flux measurements (including tight upper limits for non-detections) that were previously hindered by host galaxy contamination. This demonstrates a clear advance over standard host subtraction techniques for probing the TDE outburst and its host environment.

\section{Application Scenarios}\label{sec:limit}
In this section, we review the key assumptions we have made in the host subtraction process, and provide recommendations for the practical application of \hostsub.

To estimate the galaxy background blended with a transient, we effectively model the ratio of the flux in the vicinity of the transient (the user defined host region) and the flux within the mask on the transient (Figure~\ref{fig:input}). This is achieved by separating the 2D spectrum as the product of a 1D spectrum ($F_\mathrm{sub}$) and a 2D spatial profile ($\xi_\mathrm{sub}$). If the host galaxy region does not contain sufficient photons, the $F_\mathrm{sub}$ extracted from this region will be very noisy and might even change sign across the wavelength range. This could introduce singularities in $\xi_\mathrm{sub}$, leading to unpredictable biases in the host-subtraction results. The seeing matching process also requires a clean galaxy spatial profile in the 2D spectrum. Consequently, \hostsub\ only applies when the host galaxy is spatially resolved, i.e., the angular size to be at least a few times greater than the seeing.

We further assume that the 2D profile $\xi_\mathrm{sub}$ varies slowly on both the spatial and spectral directions and that the galaxy profile in archival broad-band images should act as a good prior of $\xi_\mathrm{sub}$. 
However, near narrow galaxy emission lines from \ion{H}{2} regions, $\xi_\mathrm{sub}$ could fluctuate abruptly along the spectral direction, as the spatial distribution of \ion{H}{2} regions (producing emission lines) is often different from that of the older stellar population (dominating the continuum). For most cases in the local universe, the broad-band fluxes, and thus our priors on the 2D galaxy profile, are dominated by the older stellar populations. As a result, $\xi_\mathrm{sub}$ typically deviates significantly from the prior derived from broad-band images, so there is no guarantee that the adopted prior is better than a non-informative one. Nevertheless, by assigning finer batches to the emission lines {(Section~\ref{sec:pipeline_host_lines})} and the customized covariance kernel {(Section~\ref{sec:pipeline_main})}, the $\xi_\mathrm{sub}$ in the narrow line regions will not affect the modeling of $\xi_\mathrm{sub}$ elsewhere. 

{We assume that the galaxy's spectral line features remain at fixed wavelengths along the slit, effectively neglecting any radial velocity dispersion within the host galaxy when modeling the 2D spectrum. This assumption is valid when the velocity gradient along the slit is small compared to the spectral resolution. However, for rapidly rotating disk galaxies, it may be necessary to adjust the spectral resolution in the GP model to account for the line tilting in the 2D spectrum.}

Another case where archival images fail to approximate the galaxy light profile in the 2D spectrum is when the transient overlaps a variable source, such as an active galactic nucleus. Additionally, when the wavelength range in the spectrum exceeds the reference image coverage {(see Figure~\ref{fig:host_prior})}, the prior function must be extrapolated beyond the photometric coverage, in which case it may be less reliable. 

Once the primary assumptions are satisfied -- namely, when the host background of the transient is spatially resolved, non-variable, and the archival images provide a complete wavelength coverage -- \hostsub\ can be universally applied to all long-slit observations of transients, irrespective to the transient types, phases, locations, or the host galaxy properties, in which cases it should provide a more robust host contamination removal than classic methods.


\section{Summary and Conclusions}\label{sec:conclusion}
We have presented a novel technique for precise host galaxy subtraction in the long-slit spectroscopy of extragalactic transients. We extract knowledge of the galaxy light distribution from multi-band reference images in archival imaging surveys, and inject the knowledge into the 2D spectrum modeling process with GPs. When the host galaxy is spatially resolved in the 2D spectrum, this GP method robustly evaluates the galaxy continuum blended with the transient spectrum. For transients emerging at complex host environments, it consistently outperforms the classic host subtraction methods, which assumes the galaxy background is an interpolation (with a simple analytical form, e.g., a linear or B-spline interpolation) of the local galaxy flux centered at the transient's trace, on both the synthetic data and real long-slit observation.

Using the software implementation of this GP method, \hostsub\ (available at \url{https://github.com/slowdivePTG/HostSub_GP}), we have substantially improved the data quality of the archival Keck/LRIS spectra for (i) a nebular-phase SN, SN\,2019eix, and (ii) a nuclear transient, AT\,2019qiz. The re-reduced spectrum of SN\,2019eix provides smoking-gun evidence on its nature as the thermonuclear explosion of a sub-{Chandrasekhar-mass} WD. For AT\,2019qiz, we set tighter constraints on the delayed emergence of narrow nebular emission (\ion{He}{2} and H$\beta$) and coronal emission lines (highly ionized Fe). 

The utility of \hostsub\ is not limited to these two specific cases -- instead it can be applied to all long-slit observations of transients with a well resolved, stationary host background. Other examples in the literature include the extraction of the late-time ($+$1200\,d) spectrum of SN\,2021aefx obtained with VLT/FORS2, {in which a clean host subtraction reveals} that the SN emission is dominated by a light echo \citep{21aefx_2025}. {\hostsub\ has also been applied to the real-time monitoring of AT\,2025ulz \citep{Kasliwal_2025}, the potential electromagnetic counterpart of the sub-threshold gravitational wave trigger S250818k \citep{2025GCN.41437....1L}. By removing the strong galaxy background, \hostsub\ reveals a hydrogen P-Cygni profile in AT\,2025ulz, which links the transient to an SN origin.}

Finally, designed to be run in a semi-automated fashion, \hostsub\ can be naturally implemented in the data reduction and classification pipeline for new transients. Developed upon \texttt{JAX} and \texttt{tinygp}, \hostsub\ enables efficient data-reduction on a personal laptop within a couple of minutes. Future implementation of GPU acceleration under the \texttt{JAX} framework will allow near real-time reduction of large spectroscopic datasets.


\software{\texttt{Astropy} \citep{Astropy_2013, Astropy_2018}, \texttt{JAX} \citep{jax_2018}, \texttt{JAXopt} \citep{jaxopt_implicit_diff}, \texttt{tinygp} \citep{tinygp_2024}, \pypeit\ \citep{pypeit:joss_pub}}

\begin{acknowledgments}
{We thank the anonymous referee for the comments that significantly improved this paper.}
We thank Erik Tollerud, J. Xavier Prochaska, Brad Holden, Kyle Westfall, Ryan Foley, Greg Aldering, and Peter Nugent for fruitful discussions. C.L. and~A.A.M.~are supported by DoE award \#\,DE-SC0025599, while A.A.M.~is also supported by Cottrell Scholar Award \#\,CS-CSA-2025-059 from Research Corporation for Science Advancement.

This work used computing resources provided by Northwestern University and the Center for Interdisciplinary Exploration and Research in Astrophysics (CIERA). This research was supported in part through the computational resources and staff contributions provided for the Quest high performance computing facility at Northwestern University which is jointly supported by the Office of the Provost, the Office for Research, and Northwestern University Information Technology.

This research has made use of the Keck Observatory Archive (KOA), which is operated by the W. M. Keck Observatory and the NASA Exoplanet Science Institute (NExScI), under contract with the National Aeronautics and Space Administration.
\end{acknowledgments}

\bibliography{main,software,facility}{}
\bibliographystyle{aasjournalv7}



\end{CJK*}
\end{document}